\documentclass[prb,twocolumn,showpacs,a4]{revtex4}

\usepackage{graphicx}
\usepackage{color}
\usepackage{amsmath,amssymb}
\usepackage{hyperref}

\definecolor{brown}{rgb}{0.6,0.3,0}
\definecolor{purple}{rgb}{0.3,0.0,0.6}

\begin{document}

\title{Kondo physics and orbital degeneracy interact to boost thermoelectrics \\ on the nanoscale}

\author{J. Azema, A.-M. Dar\'e, S. Sch\"afer, P. Lombardo}\email{pierre.lombardo@univ-amu.fr}
\address{Aix-Marseille Univ and Institut Mat\'eriaux Micro\'electronique et Nanosciences de Provence (IM2NP -- UMR 7334 CNRS)\\
Facult\'e des Sciences de St J\'er\^ome, 13397 Marseille, France}

\date{June 20, 2012}

\begin{abstract}
We investigate the transport through a nanoscale device consisting of a degenerate double-orbital Anderson dot coupled to two uncorrelated leads. We determine the thermoelectric transport properties close to the one-electron regime and compare them to a corresponding single-orbital dot. The linear and nonlinear regimes are addressed, the latter via a non-equilibrium generalization of the non-crossing approximation based on the Keldysh formalism. 
Power output and efficiency in the Kondo regime are shown to be strongly enhanced through the presence of a second orbital. 
We predict an experimentally relevant optimal operating point which benefits from the concomitant increase of the Kondo temperature in the two-orbital setup.
An approximation based on the transport coefficients and fulfilling the thermodynamic balance is proven to remain appropriate even far beyond the expected range of validity of such approaches.
Finally, the double-orbital Kondo regime reveals itself as a promising candidate to avoid, at least partially, the generic dilemma between optimal thermoelectric efficiency on one hand, and fair power output on the other.
\end{abstract}
\maketitle

\section{Introduction.}

Over the last decade, the thermoelectric effects in nanoscale devices have become a field of intense research~\cite{DD2011}, driven by the desire to  reconvert waste heat back into usable electric energy.   In addition to the obvious benefits for applications in microelectronics, where increasing clock rates and circuit integration make waste heat evacuation and power consumption serious issues, the subject is also interesting from a purely fundamental point of view: due to the high, essentially quantum-dot-like confinement in the conduction channel, electronic interactions and many-body correlations are expected to govern the transport properties of nanoscale devices. The most prominent effect is the so-called Coulomb blockade, which tends to deteriorate the electric conductance through the conduction channel.  Conversely, the Kondo effect, arising from resonant spin-flip scattering between the conduction electrons in the leads and electrons confined in the  dot-like channel, allows to overcome the Coulomb blockade at low temperatures, thereby restoring the device's electric conductance~\cite{Getal1998,COK1998, WFF2000}.

The impact of Kondo physics on the thermoelectric properties has become a topic under active experimental \cite{GSL1999,SBR2005} and theoretical \cite{BF01,DL2002,KH2002,KK2006,MMM2008,NKK2010,LSX2010,WS2010,CZ2010,RZ2012} investigation.  We would like to highlight the beautiful experiments on single-walled carbon nanotubes in the presence of transition metal impurities \cite{GSL1999} where a giant thermopower has been revealed; and on lateral quantum dots where Kondo correlations in combination with electron-hole asymmetry have been found to strongly enhance the thermopower signal \cite{SBR2005}. 

Recent theoretical works~\cite{LWF2010,NXL2010, SB2011,KLK2011,WS2012,AOG2012,MG2012} have been dedicated to thermopower generation by nanoscale devices, focusing on efficiency and power output outside the Kondo regime. Mahan and Sofo \cite{MS96} predicted earlier that large values of the thermoelectric figure of merit $ZT$ require a transmission function of vanishing width across the device which, in turn, implies a very weak coupling between localized channel state(s) and leads, and ultimately results in low conductance and power output.  A similar trade-off was uncovered by Nakpathomkun {\it et al.} \cite{NXL2010} in low dimensional systems.

In this paper, we advocate that the Kondo effect may alleviate this dilemma for a quantum-dot like conduction channel: its Abrikosov-Suhl-Kondo (ASK) resonance can be narrowed at will by rising the on-dot Coulomb interaction $U$, no matter how large the value of the lead-dot hybridization $\Gamma$ is. This route seems viable provided that temperature and voltage-bias effects do not erode the underlying Kondo physics. Formerly considered as an almost exclusive low-temperature phenomenon, the last decade's experimental progress boosted the Kondo effect to temperature ranges as high as several hundreds of Kelvin \cite{CRSL2012} and produced the first functional nanodevices with Kondo temperatures around $30\,{\rm K}$ \cite{LSBLP2002,PCHFAR2007}. Improvements can also be expected from the degenerate two-orbital Kondo effect, where entangled orbital and spin degrees of freedom have been shown, experimentally \cite{Expmulti} and theoretically \cite{Theomulti}, to raise the Kondo temperature significantly with respect to the standard spin-$\frac{1}{2}$ Kondo effect. 

In the following, we calculate thermal and electric currents through a doubly-degenerate quantum dot.  The calculation is performed in a fully nonlinear manner within the framework of the Keldysh Green's function formalism.  Specifically, we use an out-of-equilibrium generalization of the non-crossing approximation (NCA) \cite{WM94,HKH98}, which enables the computation of thermoelectric efficiency and power output, in presence of finite voltage and temperature biases applied across the quantum dot.  The benefits for the thermoelectric transport properties in the two-orbital Kondo regime become obvious in comparison to the standard single-orbital Kondo effect. Finally, we suggest an experimentally relevant optimal operating point.

\section{Model and method.}
The device considered in the following consists of a central dot coupled to two uncorrelated leads.  At simplest level, retaining only one or two relevant orbitals in the dot, the situation is readily described by an Anderson model, with a Hamiltonian given in standard notation by \cite{A61}
\begin{eqnarray}
\label{Hamiltonian}
H&=&
\epsilon_{0} \sum_{m,\sigma} c_{m \sigma }^{\dagger } c_{m \sigma } 
\,+\,\frac{U}{2}\sum_{{(m,\sigma)\neq}\atop {(m^\prime,\sigma^\prime})} n_{m\sigma} n_{m^\prime \sigma^\prime }
\nonumber  \\
&+& \sum\limits_{{\alpha\in\{L,R\}}\atop {k,m,\sigma}}\epsilon_{\alpha k}\,a_{\alpha k m\sigma }^{\dagger} a_{\alpha k m\sigma}
\nonumber \\
&+& \sum\limits_{{\alpha\in\{L,R\}}\atop {k,m,\sigma}} 
\left( t_{\alpha} c_{m\sigma }^{\dagger} a_{\alpha k m\sigma}+t_{\alpha}^*a_{\alpha k m\sigma }^{\dagger } c_{m \sigma }\right)
\,\mbox{.}
\end{eqnarray}
The first line concerns the single ($m=1$) or doubly degenerate ($m=1,2$) orbital with on-site Coulomb repulsion $U$, the second line describes the left (L) and right (R) leads, and the last line accounts for the hybridization tunneling between dot and leads which conserves orbital and spin quantum numbers. In the double-orbital case the above Hamiltonian is $SU(4)$-degenerate; nevertheless our results do not hinge on this symmetry.

The results are found to depend little on the specific shape and width of the lead densities of states (DOS), provided that they are smooth enough and that their widths are by far the largest energy scale in the problem. For the solution of the NCA equations, we may therefore use a Gaussian spin-summed  DOS, $N_{\alpha}(\varepsilon)$, characterized by the same half-width $D$ for both leads.  The $k$-independence of the tunneling amplitudes yields a hybridization between dot and leads that is directly proportional to the respective lead DOS, $\Gamma_{\alpha}(\varepsilon)=\pi t_{\alpha}^2 N_{\alpha}(\varepsilon)$, (where henceforth a symmetric coupling, $\Gamma_{\rm R}(\varepsilon)=\Gamma_{\rm L}(\varepsilon)$, will be assumed). The total hybridization strength at the {\em mean} Fermi energy,  $\overline{\mu}=\frac{1}{2}(\mu_{\rm L}+\mu_{\rm R})$, will serve as an energy unit:  $\Gamma=\Gamma_{\rm L}(\overline{\mu})+\Gamma_{\rm R}(\overline{\mu})$.

As appropriate for the power generating regime, the temperature bias $\Delta T=T_{\rm L}-T_{\rm R}>0$ across the device is applied in the opposite direction of the voltage bias $-e V_{\rm b}=\mu_{\rm L}-\mu_{\rm R}<0$.
The nonlinear electrical and hot (left) lead thermal currents are given, respectively, by
\begin{eqnarray}
I&=&\frac{2e}{h}\int \left[f_{\rm L}(\varepsilon)-f_{\rm R}(\varepsilon)\right]\tau(\varepsilon)\,{\rm d}\varepsilon \qquad\mbox{and}
\nonumber
\\
\dot{Q_{\rm L}}&=&\frac{2}{h}\int(\varepsilon-\mu_{\rm L}) \left[f_{\rm L}(\varepsilon)-f_{\rm R}(\varepsilon)\right]\tau(\varepsilon)\,{\rm d}\varepsilon
\,\mbox{,}
\label{IQeqs}
\end{eqnarray}
where $h$ denotes Planck's constant and $-e$ the electronic charge. $f_{\alpha}(\varepsilon)\equiv f(\varepsilon-\mu_{\alpha})$ are the lead Fermi functions, and $\tau(\varepsilon)=\frac{\pi}{4} A(\varepsilon)\Gamma(\varepsilon)$, with $A(\varepsilon)=-\frac{1}{\pi}{\rm Im}[\sum_{m,\sigma}G_{m\sigma}(\varepsilon+i\delta)]$, the total dot spectral density. The retarded Green's function $G_{m\sigma}(\varepsilon+i\delta)=\langle\langle c_{m\sigma}; c_{m\sigma}^{\dagger}\rangle\rangle$ is obtained from a generalized Keldysh-based out-of-equilibrium NCA~\cite{WM94,HKH98}. The NCA is known to reliably describe the low-temperature Kondo scale $T_{\rm K}$, and to give accurate results for the dot spectral density down to temperatures of the order of a fraction of $T_{\rm K}$~\cite{H97,NCArecent}.
The power output is given by the sum of the heat currents, $P=\dot{Q}_L+\dot{Q}_R$, which together with Eqs.~(\ref{IQeqs}) yields $P=I V_{\rm b}$. The thermoelectric converter efficiency $\eta=P/\dot{Q_{\rm L}}$ is conveniently normalized in terms of the Carnot efficiency, $\eta_{\rm C}=1-T_{\rm R}/T_{\rm L}$. In the following, we focus on power generation, {\sl i.e.} positive $I$.

The transport in the {\em linear} regime is obtained by a first order series expansion of the Fermi functions in Eqs.~(\ref{IQeqs}) characterized by the following transport coefficients: electrical conductance $G(T)$, Seebeck coefficient $S(T)$, and electronic thermal conductance $K_{\rm e}(T)$,
\begin{subequations}
\label{GSKeqs}
\begin{eqnarray}
G(T) &=& e^{2} I_{0}(T) \,\mbox{,} \\
S(T) &=& -\frac{1}{eT}\frac{I_{1}(T)-\mu I_{0}(T)}{I_{0}(T)} \,\mbox{,} \\
K_{\rm e}(T) &=& \frac{1}{T}\left[I_{2}(T) - \frac{I_{1}^{2}(T)}{I_{0}(T)} \right] 
\,\mbox{,}
\end{eqnarray}
\end{subequations}
given  in terms of the transport integrals $I_{n}$ (defined for $n=0,1,2$) \cite{KH2002}
\begin{equation}
\label{Ineq}
I_{n}(T) = \frac{2}{h}\int {\rm d}\varepsilon\,\varepsilon^{n}\,\tau^\mathrm{eq}(\varepsilon)\left(-\frac{\partial f}{\partial \varepsilon}\right)
\,\mbox{,}
\end{equation}
where $\tau^\mathrm{eq}(\varepsilon)=\frac{\pi}{4} A^\mathrm{eq}(\varepsilon)\Gamma(\varepsilon)$ has the role of an equilibrium transmission function.
Finally, retaining only the electronic contributions to the thermal conductance, the figure of merit is defined by $ZT=S^2GT/K_{\rm e}$.

\section{Results.}
Taking $\overline{\mu}$ as the energy origin, the chosen parameters $\varepsilon_{0}=-3.2\Gamma$ and $U=16\Gamma$, are close to the one-electron Kondo regime, implying approximately half filling for a single-orbital (SO), and quarter filling for the degenerate double-orbital (DO) case. More precisely, in the explored temperature range, the average number of electrons on the dot increases with $T$, and lies in the interval $[ 0.86- 0.9 ]$ for SO, and $[ 0.76- 0.9 ]$ for DO.
\begin{figure}
\begin{center}
    \includegraphics[width=8cm]{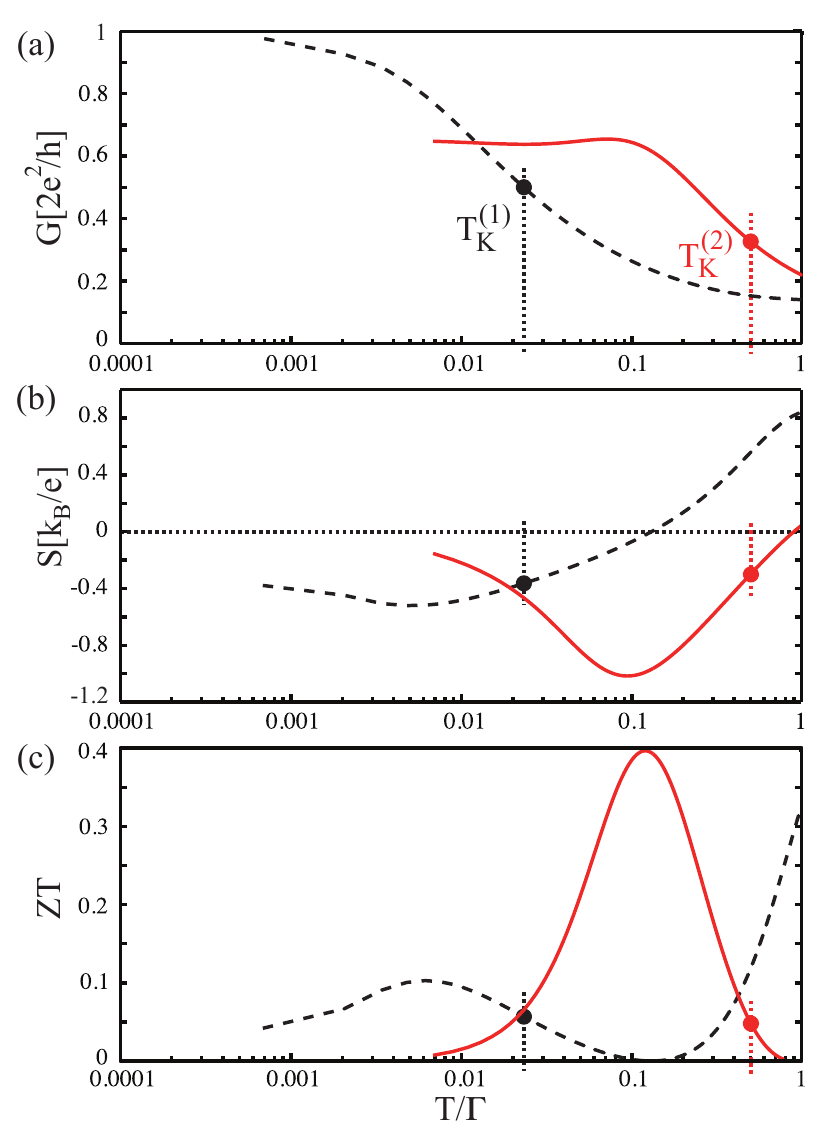}
    \caption{(a) Electric conductance $G(T)$, (b) Seebeck coefficient $S(T)$, and (c) figure of merit $ZT$ vs. $T/\Gamma$, for $\varepsilon_{0}=-3.2\Gamma$ and $U=16\Gamma$.  Dashed lines refer to the SO and plain red lines to the DO cases. Vertical dotted lines indicate the respective Kondo temperatures.\label{fig1}}
\end{center}
\end{figure}

\subsection{Linear transport properties.}
First we discuss the linear thermoelectric transport properties, some of which confirm earlier results by Sakano {\sl et al.} \cite{SKK2007}: in Fig.~\ref{fig1} the temperature dependences of $G$, $S$ and $ZT$ are plotted for the SO and DO cases~\cite{rem1}. In the low-temperature regime, the Kondo effect allows to overcome the Coulomb blockade due to the presence of a robust ASK resonance in the immediate vicinity of the Fermi level which provides the necessary spectral density and thus guarantees conductance.
Defining the Kondo temperature as the value where $G(T)$ reaches half its maximum, we find $T^{(1)}_{\rm K}\approx 0.023\Gamma$ for the SO case (with $k_{\rm B}=1$ henceforth), and a much larger $T^{(2)}_{\rm K}\approx 0.5\Gamma$ for the DO case. Such an enhancement is in line with e.g. Hewson's textbook \cite{H97} and should be relevant for potential applications.
As obvious from the dashed curve in  Fig.~\ref{fig1}a, the low-temperature conductance for the SO case saturates close to the spin-summed quantum of conductance $2e^2/h$, implying that the conductance contribution of each spin species is almost maximal, as expected for a device close to half filling.  This contrasts with the corresponding low-temperature conductance for the DO case (full red line): for $T\ll T^{(2)}_{\rm K}$, we observe a large plateau where the conductance is almost temperature-independent.
The height of the plateau does not quite reach $2e^2/h$ expected at quarter filling, but saturates at $G \sim 1.2 e^2/h$, as predicted for an occupancy $\langle n_{m \sigma}\rangle  \sim 0.19$, by the Landauer formula in combination with the Friedel sum rule: $G=2e^2/h \sum_{m}\sin^2(\pi \langle n_{m\sigma}\rangle)$ (see e.g. Ref.~\cite{SK06}).
The DO conductance plateau ends to its right with a soft bump whose origin follows directly from the behavior of the spectral density: the ASK resonance is found to lie slightly beyond the Fermi energy, such that thermal broadening effects first enhance the spectral density at the Fermi level, while only larger temperature rises yield to the inevitable erosion of conductance.

Fig.~\ref{fig1}b displays the thermoelectric power as a function of $T$, whose sign indicates the particle- or hole-like nature of the transport~\cite{BF01}. Again, the Kondo effect is responsible for the decisive feature, {\sl i.e.} the low-temperature minimum of the Seebeck coefficient: due to its narrowness and location right above the Fermi level, the ASK resonance is very efficient in capturing a fraction of the hottest electrons of the left electrode, thereby giving rise to the thermoelectric effect.  Approaching $T_{\rm K}$ the Kondo resonance vanishes and the effect disappears. Upon a further temperature rise, contributions from occupied states dominate which results in a sign change in $S$. For even higher temperatures, the unoccupied states with energies around $\varepsilon_{0}+U$ reverse the sign change (off range).
As can be seen from Fig.~\ref{fig1}b, this negative extremum in the Seebeck coefficient is located at roughly a quarter of the Kondo temperature.  With respect to the SO case, the DO minimum is not only shifted to higher temperatures, but also benefits from a Kondo resonance with significantly enhanced  spectral weight, yielding a minimum which is  almost twice as deep as the SO minimum. Although the present interpretation is strictly speaking only appropriate for the linear regime, it is confirmed in its main lines by the fully nonlinear calculation presented later in this section.

As a direct result of the aforementioned behavior of the Seebeck coefficient, the figure of merit, shown in Fig.~\ref{fig1}c, benefits from this situation: 
its maximum is strongly enhanced in the DO case, with the maximum $ZT\approx 0.4$ occurring for $T^{(2)}_{\rm opt}\approx 0.25 T^{(2)}_{\rm K}\approx 0.12\,\Gamma$, {\sl i.e} well above the SO maximum $ZT\approx 0.1$ at $T^{(1)}_{\rm opt}\approx 0.006\,\Gamma$.
Parts of this enhancement are also due to the dip occurring in the ratio $K_{\rm e}/(GT)$ for temperatures close to $T_{\rm K}$, indicating a violation of the Wiedemann-Franz law, as noticed earlier by Krawiec {\it et al.}~\cite{KW2007}.
Although the temperature for which the figure of merit is maximal yields roughly the optimal efficiency of  thermoelectric devices, we want to stress that this criterion has to be verified beyond the linear framework.

\subsection{Nonlinear transport.}
\begin{figure}
\begin{center}
    \includegraphics[width=8.5cm]{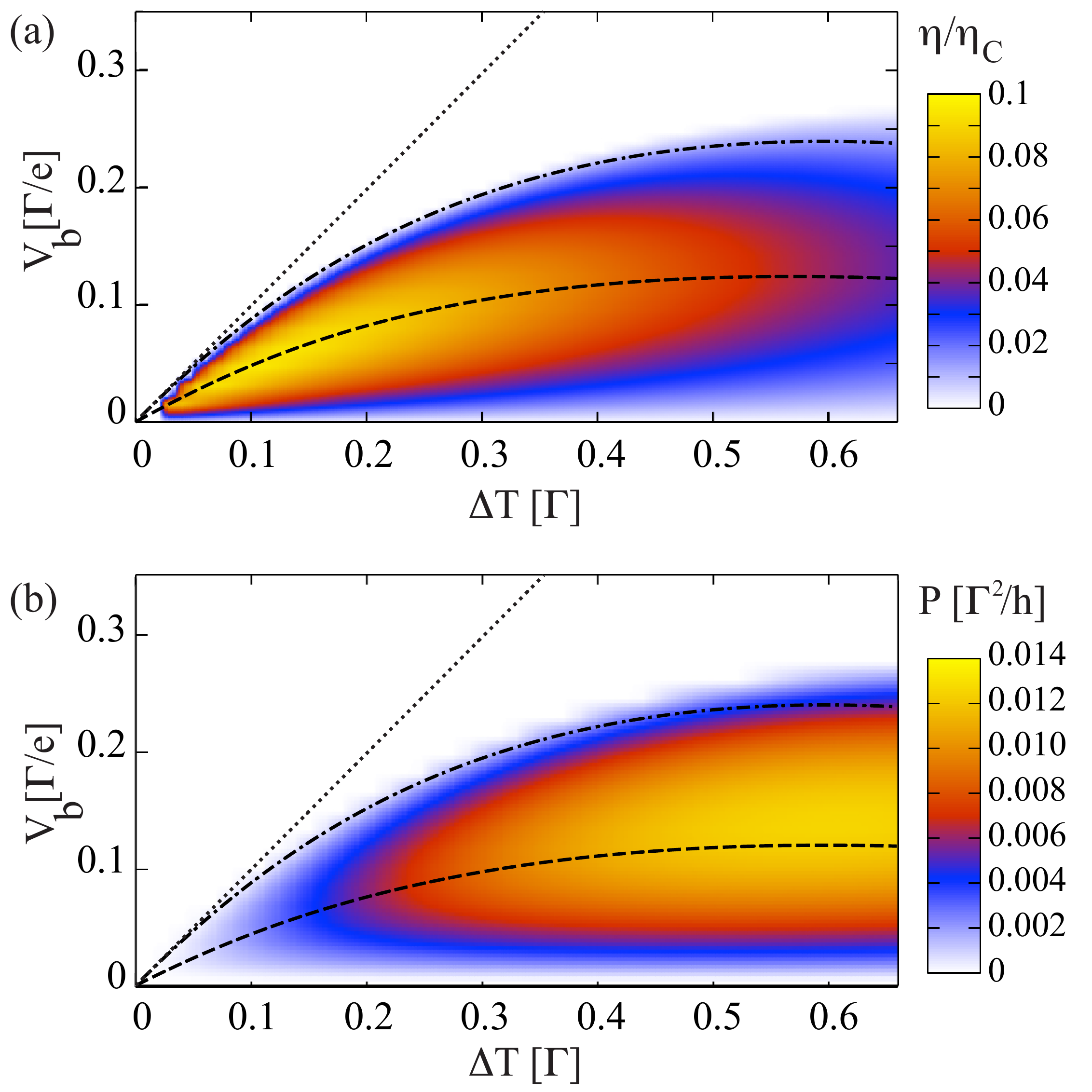}
    \caption{Nonlinear transport for the DO case in the power generating regime for $\varepsilon_{0}=-3.2\,\Gamma$, $U=16\,\Gamma$, and fixed $T_{\rm R}=0.12\,\Gamma$ as functions of voltage and temperature biases across the device. (a) Normalized efficiency $\eta/\eta_{\rm C}$. (b) Power output. 
The various lines correspond to approximations based on transport coefficients: 
the dot-dashed and dashed lines represent, respectively, the lines of zero power output and the ridge lines for efficiency and power output for the TCA; the straight dotted line corresponds to zero power output of prosaic linear response theory (see main text).
\label{fig2}}
\end{center}
\end{figure}
The results of such a fully nonlinear calculation for the DO setup are illustrated in Fig.~\ref{fig2}, displaying the transport properties as functions of $V_{\rm b}$ and  $\Delta T$, which is applied such that the cold (right) lead is at constant temperature, $T_{\rm R}\approx T^{(2)}_{\rm opt}$.
As obvious from the figure, efficiency and power output both exhibit a maximum, albeit for different regimes: the maximal efficiency, closely related to the ASK resonance, is reached for $\Delta T\sim T_{\rm R}$, while the power output is maximal for (probably experimentally unachievable) large $\Delta T\sim 5T_{\rm R}$ where the ASK resonance is strongly suppressed.
Our calculations predict an optimal efficiency of 10\% of the Carnot efficiency for $V_{\rm b}\approx 0.05\Gamma/e$ and $\Delta T\approx 0.1\Gamma$. 
Although the power output is not yet maximal at optimal efficiency, it reaches appreciable values of the order of $2.13\ 10^{-3}\,\Gamma^2/h$ which could be interesting for potential applications.
For an order-of-magnitude estimate, we assume a hybridization $\Gamma = 25\,{\rm meV}$, yielding a power output of $52\,{\rm pW}$ at maximum efficiency which is reached for $V_{\rm b}=1.2\,{\rm mV}$, $T_{\rm R}=35\,{\rm K}$ and $T_{\rm L}=64\,{\rm K}$.

The latter results should be compared to a Coulomb-blockade regime of similar efficiency.  Postulating the existence of an adjustable Lorentzian peak in the spectral function and choosing its width and position such that the efficiency maximum matches our calculations -- {\sl i.e.} 10\% of the Carnot efficiency -- we find that the power output in the DO Kondo regime surpasses that of the Coulomb blockade regime by about $30\%$. 
This increase constitutes a significant step towards the resolution of the dilemma between optimal efficiency and power. For analogous reasons to those mentioned in the linear regime, the physical origin of the power increase in the Kondo regime is due to the narrowness and position of the ASK resonance. We stress that this feature is not eroded by the present moderate voltage biases~\cite{WM94}.
This clearly contrasts with the Coulomb blockade regime, where the width of the resonance is directly proportional to the hybridization, resulting in the aforementioned dilemma between large $\Gamma$, necessary for large power output, and small $\Gamma$, required for efficiency.
From the experimental point of view, another advantage is that the ASK resonance is automatically anchored in the vicinity of the lead Fermi levels, and hence does not require any fine-tuning of the level position $\varepsilon_{0}$.  As a result, the thermogenerator remains operational for $\varepsilon_{0}$ varying from $-4.4\Gamma $ to $0.7\Gamma$, {\sl i.e.} a range three times wider than in the pure Coulomb blockade regime.
Furthermore, in line with our expectations, the DO setup clearly outperforms the standard SO one.  In the latter case, the maximum efficiency is at only $3\%$ of the Carnot efficiency, occurring at $V_{\rm b}\approx 2.38\ 10^{-3}\,\Gamma/e$ and $\Delta T \approx 9.6\ 10^{-3}\,\Gamma$, and the SO power output does not exceed $6.62\ 10^{-6} \,\Gamma^2/h$. 

Finally, we want to address the question whether similar results could have been obtained within the framework of an approximation which solely uses the transport coefficients (TCA).  These coefficients were calculated in the previous section and depend only on the device's equilibrium properties. To this end~\cite{HU1961}, we calculate the associated power output $P^{\rm TCA}=V_{b}I^{\rm TCA}=-G V_{b}^2-SGV_{b}\Delta T$, and heat currents $\dot Q_{L}^{\rm TCA}=SGT_{L}V_{b}+K_{e}(1+ZT_{L})\Delta T-\frac{1}{2}[I^{\rm TCA}]^2/G$ and $\dot Q_{R}^{\rm TCA}=-SGT_{R}V_{b}-K_{e}(1+ZT_{R})\Delta T-\frac{1}{2}[I^{\rm TCA}]^2/G$. 
The transport coefficients in the latter expressions, $S$, $G$, $K_{e}$ and $Z$, are evaluated~\cite{MG2012} at the device operating temperature $\overline T = \frac{1}{2}(T_{\rm L}+T_{\rm R})$.
Note that each reservoir temperature is used for the heat flux leaving the corresponding electrode~\cite{HU1961}. 
The above approximation also accounts for the Joule heating of the device (which is small in our case), and presents the virtue of fulfilling the thermodynamic balance $P=\dot{Q}_L+\dot{Q}_R$.
A detailed  comparison reveals that this approximation reproduces the results of the fully nonlinear approach, qualitatively on the whole plot range of Fig.~\ref{fig2}, while small quantitative discrepancies appear for $\Delta T\gtrsim 4 T_{\rm R}$.
This remarkable agreement is illustrated in the figure by overlaying, on the fully nonlinear results, the TCA lines of zero power output (dot-dashed) and the TCA ridge lines (dashed).
The efficiency ridge line is given~\cite{HU1961} by $\eta_{\rm max}^{\rm TCA}/\eta_{C}=(\sqrt{1+Z\overline{T}}-1)/(\sqrt{1+Z\overline{T}}+T_{\rm R}/T_{\rm L})$.
At this point, it is worth mentioning that linear response theory implemented around one of the reservoir temperatures -- in this case $T_{\rm R}$ -- yields~\cite{note} the straight dotted line of zero power output in Fig.~\ref{fig2} which rapidly departs from the fully nonlinear results.
This situation lead us to the conclusion that the discrepancy between such a prosaic linear response theory and the fully nonlinear results does not necessarily require a genuine out-of-equilibrium approach -- even if this is sometimes claimed~\cite{LWF2010,MG2012} -- and, conversely, that carefully implemented approximations  relying on transport coefficients should be able to reproduce the nonlinear results rather accurately for similar devices.


\section{Conclusion.}
In this paper, we have studied the thermoelectric transport through a double-orbital quantum dot in the Kondo regime. We showed that transport properties and power generation are strongly enhanced by the interplay of orbital degeneracy and Kondo physics. This setup outperforms not only the usual single-orbital quantum dot, but also devices in the Coulomb blockade regime.  An experimentally relevant optimal operating point has been determined. It owes its stability to the Kondo effect and might even allow for a complete removal of the gate electrode.  
A comparison of the transport properties obtained from carefully implemented approximations relying solely on the transport coefficients
with the fully nonlinear solution proves the robustness of the former up to large applied temperature bias for the systems under study, thus adding credibility to many works on nanoscale devices conducted within this framework.


\begin{thebibliography}{100}
\bibitem{note} With the power output approximated by   
%
$-GV_{b}^2-SGV_{b}\Delta T$ 
where the transport coefficients are evaluated at $T_{R}$.
\end{thebibliography}
\end{document}